\documentclass[11pt]{iopart}
\pdfoutput=1
\usepackage{epsfig,graphicx,latexsym,iopams}
\usepackage{color}
\newcommand\beq{\begin{equation}}
\newcommand\eeq{\end{equation}}
\newcommand{\ba}{\begin{eqnarray}}
\newcommand{\ea}{\end{eqnarray}}  
\def\spose#1{\hbox to 0pt{#1\hss}}
\def\lta{\mathrel{\spose{\lower 3pt\hbox{$\mathchar"218$}}
     \raise 2.0pt\hbox{$\mathchar"13C$}}}
\def\gta{\mathrel{\spose{\lower 3pt\hbox{$\mathchar"218$}}
     \raise 2.0pt\hbox{$\mathchar"13E$}}}

\begin{document}

\title{Constraining properties of the black hole population using LISA}

\author{
  Jonathan R. Gair$^1$\footnote[3]{email: jgair@ast.cam.ac.uk},
  Alberto Sesana$^2$,
  Emanuele Berti$^{3,4}$,
  Marta Volonteri$^5$
}

\address{$^1$Institute of Astronomy, University of Cambridge, Cambridge, CB3 0HA, UK}
\address{$^2$Albert Einstein Institute, Am Muhlenberg 1 D-14476 Golm, Germany}
\address{$^3$Department of Physics and Astronomy, The University of Mississippi,
  University, MS 38677-1848, USA}
\address{$^4$California Institute of Technology, Pasadena, CA 91109, USA}
\address{$^5$Department of Astronomy, University of Michigan, Ann Arbor, MI, USA}

\begin{abstract}
LISA will detect gravitational waves from tens to hundreds of systems
containing black holes with mass in the range
$10^4M_{\odot}$--$10^7M_{\odot}$. Black holes in this mass range are not well
constrained by current electromagnetic observations, so LISA could
significantly enhance our understanding of the astrophysics of such
systems. In this paper, we describe a framework for combining LISA
observations to make statements about massive black hole populations. We
summarise the constraints that LISA observations of extreme-mass-ratio
inspirals might be able to place on the mass function of black holes in the
LISA range. We also describe how LISA observations can be used to choose
between different models for the hierarchical growth of structure in the early
Universe. We consider four models that differ in their prescription for the
initial mass distribution of black hole seeds, and in the efficiency of
accretion onto the black holes. We show that with as little as 3 months of
LISA data we can clearly distinguish between these models, even under
relatively pessimistic assumptions about the performance of the detector and
our knowledge of the gravitational waveforms.
\end{abstract}

\section{Introduction}
Current measurements of black hole (BH) masses are almost exclusively for
systems with mass above $\sim10^6M_{\odot}$. The shape of the mass function
for less massive BHs is expected to retain a signature of the initial mass
distribution of the seeds from which BHs grow, which is erased at the high
mass end by the effects of accretion~\cite{volonteri2008}. In addition, it is
not known if the tight correlations observed between the properties of high
mass BHs and their host galaxies extend down to lower
masses~\cite{Greene2008}, which has important consequences for our
understanding of the co-evolution of galaxies and BHs. Probing BHs in the mass
range $10^4M_{\odot} < M < 10^7M_{\odot}$ is thus crucial to our understanding
of the growth of structure, and LISA~\cite{lisappa} is one of the few
instruments that has the potential to observe such systems. LISA is expected
to see a few tens of massive BH mergers (MBHMs) per year~\cite{sesana07,lisape} and as
many as several hundred extreme-mass-ratio inspirals (EMRIs) of stellar-mass
compact objects into massive BHs in the centres of
galaxies~\cite{gairEMRIAstro}. The MBHMs can be observed throughout the
Universe, while the EMRIs will only be seen at low redshift, $z\lesssim1$, but
LISA will be able to measure the parameters of both kinds of event to
unprecedented precision~\cite{lisape,AK,HG09}.

The mass function of BHs in the LISA range is uncertain due to the lack of
direct observations. If the BH population traces the active galaxy population,
then the mass function should turn over for $M \lesssim
3\times10^6M_{\odot}$~\cite{GreeneHo}. However, if the L--$\sigma$ and
M--$\sigma$ relations derived for more massive galaxies can be extrapolated to
lower masses, the observed galaxy luminosity function would imply a flat BH
mass function in this range~\cite{GreeneHo}. Applying corrections to Sloan
Digital Sky Survey measurements of the velocity dispersion instead yields a
mass function that increases toward lower masses~\cite{shethSDSS}. Overall the
present uncertainty in the slope of the quiescent BH mass function for
$M\lesssim10^7M_{\odot}$ is at least $\pm 0.3$~\cite{emrimodsel}. LISA EMRI
observations could therefore play an important role in pinning down this slope
in the low-redshift Universe.

LISA MBHM events will occur following mergers of the host galaxies of the BHs,
and thus trace the hierarchical growth of structure. Models of structure
formation are tuned to match existing observations, and therefore make similar
predictions for mergers at the high mass end of the BH mass function, but
differ significantly for lower masses. In particular, both ``light
seed''~\cite{vhm} and ``heavy seed''~\cite{bvr} models are consistent with
existing data. It is unlikely that observations in the electromagnetic
spectrum will rule out either class of models in the next decade, so LISA
could make important contributions to our understanding of the early epoch of
galaxy formation.

LISA will be able to make very precise measurements of the parameters of
individual EMRI and MBHM systems that are
observed~\cite{AK,HG09,lisape}. Precise measurements for single systems are
very important for fundamental physics~\cite{astrogr}, but it is the full set
of events that are seen which will carry the most important information for
astrophysics. In this paper, we describe a method to combine this information
in order to make astrophysical statements, which is based on a Bayesian
framework using a parametric model for the probability distribution of
observed events. LISA model selection using MBHMs was also considered
in~\cite{plowmanSMBH2} using the non-parametric Kolmogorov-Smirnov test to
compare parameter distributions. While our conclusions are broadly similar,
the framework presented here is more general and can be easily extended to
other problems.

This paper is organised as follows. In Section~\ref{methods} we describe our
approach to using LISA to place constraints on astrophysical models. This
includes a discussion of the statistical framework used for model selection
and a description of how we model LISA instrumental effects, i.e., the
completeness of LISA observations and the parameter estimation errors that
arise from noise in the detector. In Section~\ref{emri} we summarise the
constraints that LISA EMRI observations could place on the shape of the BH
mass function in the LISA range. These results were previously described
in~\cite{emrimodsel} but we briefly review them here for completeness. In
Section~\ref{smbh} we describe how LISA can be used to choose between
different models for the hierarchical growth of structure. We illustrate this
using four different models that differ in their seed mass distributions and
accretion prescriptions. These results are new and appear here for the first
time. In Section~\ref{discuss} we discuss our results and possible extensions
of this work.

\section{Methods}
\label{methods}

Our aim is to make inferences about the population of BHs in the Universe
based on LISA observations. Bayes theorem relates the posterior probability,
$p(\vec{\lambda}|D,M)$, for the parameters $\vec{\lambda}$ of a model $M$
given observed data $D$, to the likelihood $p(D|\vec{\lambda},M)$ of seeing
that data under model $M$ with parameters $\vec{\lambda}$, and the prior,
$\pi(\vec{\lambda})$ for the parameters $\vec{\lambda}$:
\begin{equation}
p(\vec{\lambda}|D,M) = \frac{p(D|\vec{\lambda},M) \, \pi(\vec{\lambda})}{{\cal Z}}, \qquad \mbox{where } {\cal Z} = \int p(D|\vec{\lambda},M) \, \pi(\vec{\lambda}) {\rm d}^N \lambda
\label{bayes}
\end{equation}
In this context, the model $M$ is a description of the population of BHs in
the Universe, the data are the parameters of the sources that LISA detects and
the uncertainty in the distributions comes from the fact that mergers occur
stochastically in the Universe --- a given model will predict the {\it rate}
at which LISA events occur but cannot predict the exact systems that LISA will
observe.

To compute the likelihood, $p(D|\vec{\lambda},M)$, we can imagine dividing the
parameter space of possible signals into bins, labelled by $i$. The data, $D$,
is the number of events, $n_i$, observed in each bin $i$. A particular model
will predict the rate, $r_i(\vec{\lambda})$, at which events in a particular
parameter bin occur in the Universe. As the events start at random times, the
number of events occurring in a given bin during the LISA mission will be
drawn from a Poisson probability distribution with rate
$r_i(\vec{\lambda})$. Events in different bins are independent and so if we
temporarily ignore LISA selection effects and parameter estimation errors, the
likelihood of seeing the set of events $D=\{n_i\}$, is
\begin{equation}
p(D|\vec{\lambda},M) = \prod_{i=1}^K \frac{(r_i(\vec{\lambda}))^{n_i} {\rm e}^{-r_i(\vec{\lambda})}}{n_i!}
\label{like}
\end{equation}
It is possible to take the continuum limit of this expression by letting the
bin volume approach zero. The above expression then becomes a product of the
point probabilities of the events observed, normalised by the total number of
events the model predicts.

For model selection, i.e., to choose the model that provides the best
description of the observed data, we use the evidence, which is the quantity
${\cal Z}$ appearing in the denominator of Bayes Theorem,
Eq.~(\ref{bayes}). To compare models A and B, we compute the odds ratio (see, for example,~\cite{mackayinf})
\begin{equation}
O_{AB} = \frac{{\cal Z}_A \, P(A)}{{\cal Z}_B \, P(B)}
\label{odds}
\end{equation}
in which $P(X)$ denotes the prior probability assigned to model $X$. If
$O_{AB} \gg 1$ ($O_{AB} \ll 1$), model $A$ (model B) provides a much better
description of the data.

The fact that we use an imperfect detector to make the observations introduces
two complications into the analysis. Firstly, the data set is not complete ---
LISA sees only a certain subset of the events that occur during the
mission. Secondly, the presence of noise in the detector introduces errors
into the parameter estimates.

The incompleteness of the LISA observation means that only a certain fraction
of the events that occur in a given bin in the parameter space will be
detected. If the completeness is known as a function of the source parameters,
then this can be included in the likelihood computation by replacing
$r_i(\vec{\lambda})$ by the effective observed rate,
$\tilde{r}_i(\vec{\lambda}) = {\cal C}_i r_i(\vec{\lambda})$, in
Eq.~(\ref{like}), where ${\cal C}_i$ is the completeness in bin $i$. In our
analysis, we assume a simple signal-to-noise ratio (SNR) cut model for
completeness. We assume that $100\%$ of events with matched-filtering SNR
$\rho > \rho_{\rm thresh}$ are detected, and that $0\%$ of events with $\rho <
\rho_{\rm thresh}$ are detected. This is a reasonable model if $\rho_{\rm thresh}$ is set
moderately high, since if events with SNR less than $\rho_{\rm thresh}$ were actually detected, they could just be excluded from the analysis. The parameters of events with lower SNR tend to be more poorly determined and thus do not contribute much to model discrimination, so there is little change in the results when such events are excluded.

Instrumental noise leads to imperfect parameter measurement. In the analysis
of actual LISA data, we will derive a posterior probability distribution (pdf)
for the parameters of the sources. Using this, the likelihood can be obtained
by integrating the continuum version of Eq.~(\ref{like}) over the
pdf~\cite{emrimodsel}. An equivalent approach, which is more convenient for
scoping out LISA's potential, is to suppose each source is assigned to the bin
in which the maximum {\em a posteriori} probability lies, and to fold the
parameter uncertainty into the effective rate of observed events,
$\tilde{r}_i(\vec{\lambda})$, in each bin. In practice,
$\tilde{r}_i(\vec{\lambda})$ can be computed by generating a large number of
realisations of the set of events that LISA observes. For each event in each
realisation the parameter uncertainty may be modelled as a Gaussian,
$p(\vec{\lambda}) \propto \exp(-\Gamma_{ij} (\lambda^i-\lambda^i_0)
(\lambda^j-\lambda^j_0)/2)$, centred on the true parameters,
$\vec{\lambda}_0$. A fractional rate $\delta \tilde{r}_i = \int_{{\cal B}_i}
p(\vec{\lambda}) {\rm d}^N\lambda$ is assigned to each bin ${\cal B}_i$. This
is similar in principle to the LISA ``error kernel'' described
in~\cite{plowmanSMBH}. The parameter error comes primarily from detector
noise, but additional errors arise from weak lensing, which changes the
apparent luminosity distance of sources at higher redshift, and from
uncertainties in the cosmological parameters used to convert from luminosity
distance to redshift. These can be included in the redshift-redshift component
of $\Gamma^{-1}$ by writing $\Gamma^{-1}_{zz} = (\Gamma_N)^{-1}_{zz} + \Delta
z^2_{WL} + \Delta z^2_{\rm cos}$, where $N$, $WL$, cos denote the
contributions from noise, weak-lensing and cosmological parameter
uncertainty. We model the weak-lensing error using results in~\cite{wangholz}
and the cosmological error following~\cite{Berti:2005qd}:
\begin{equation}
\Delta z^2_{\rm cos}=\left(\frac{\partial D_L}{\partial z}\right)^{-2}\left[\left(\frac{\Delta D_L^2}{D_L^2}+
\frac{\Delta H_0^2}{H_0^2}\right)D_L^2+\Delta\Omega_\Lambda^2\left(\frac{\partial D_L}{\partial \Omega_\Lambda}\right)\right]\,,
\end{equation}
and we assume that $\Delta H_0/H_0 = \Delta \Omega_{\Lambda}/\Omega_{\Lambda}
= 0.01$ by the time LISA flies.

\section{Results}
\label{results}
\subsection{Extreme-mass-ratio inspirals}
\label{emri}
This preceding framework was used in~\cite{emrimodsel} to explore the
constraints that LISA EMRI observations could place on the BH mass function at
low redshift. This analysis made various assumptions --- an SNR cut of
$\rho_{\rm thresh}=30$ was used to define the completeness function; all EMRIs
were assumed to be on circular and equatorial orbits, which meant the
completeness could be determined from the {\it observable lifetime} of a
particular EMRI, as defined in~\cite{gairEMRIAstro}; parameter estimation
errors were ignored in the analysis, but included in the generation of
realisations of the LISA data; the data was taken to be measurements of the
central BH mass $M$ and source redshift $z$ only; and the scaling of the
intrinsic EMRI rate per BH was assumed to be known and given by results
in~\cite{hopman09}. Assuming a simple power-law mass function, $A_0
(M/M_*)^{\alpha_0}$, and that all BHs have spin $a=0.9$, it was found that
LISA could measure the parameters to a precision $\Delta A_0 \approx 0.5
\sqrt{10/N_{\rm obs}}$ and $\Delta \alpha_0 \approx 0.2 \sqrt{10/N_{\rm obs}}$
if $N_{\rm obs}$ events were observed. This compares very well to the current
precision, $\sim \pm 0.3$, on the slope of the BH mass function, particularly
given that these models typically predict $\sim 100$s of observable EMRI
events.

Using a redshift-dependent ansatz for the mass function, $A_0 (1+z)^{A_1}
(M/M_*)^{\alpha_0 - \alpha_1 z}$, it was found that LISA would not be able to
place reasonable constraints on the evolution parameters $A_1$,
$\alpha_1$. This is because the majority of EMRI events will be detected at
low redshift, $z \lesssim 1$. The main caveat in these results is the
assumption that the mass-dependence of the EMRI rate will be known by the time
LISA flies. These results can also be interpreted as the precision with which
the convolution of the mass function with the EMRI rate can be
determined. More work is required to determine if combined EMRI and MBHM
observations can decouple these effects and perhaps measure evolution in the
mass function.

\subsection{Comparable mass black hole mergers}
\label{smbh}
LISA observations of MBHMs can be used to choose between different models for
the growth of structure. There are various models for the hierarchical
assembly of galaxies, but these have been tuned to fit existing
electromagnetic observations which do not constrain BHs in the mass range of
interest to LISA. The models therefore make quite different predictions for
the expected set of LISA events, which means that LISA has the potential to
discriminate between them. We consider four different models, which differ in
the prescription for the masses of the initial seeds from which BHs grow and
in the prescription for accretion onto the BHs. We consider two ``light seed''
models~\cite{vhm}, in which BH seeds of mass $\sim100M_{\odot}$ form as
remnants of metal-free stars at redshift $z\gtrsim20$, and two ``heavy seed''
models~\cite{bvr}, in which seeds with mass $\sim10^5M_{\odot}$ form directly
from the collapse of massive protogalactic disks in the redshift range $10
\lesssim z \lesssim 15$. In each case, we consider two accretion prescriptions
\cite{bv}: (i) ``coherent'' accretion, in which material accreting onto the black hole tends to have similar angular momentum~\cite{bardeen,thornespin}, which could occur if the large scale structure of the feeding material is in a disc-like configuration~\cite{dotti09,dotti10}; (ii) ``chaotic'' accretion, in which there
are many short accretion episodes with different angular momentum spin axes in
each one~\cite{kingpringle}. The four models are summarised in
Table~\ref{modelsumm}. We chose these four models to allow easier comparison
to the literature. The same four models were used to explore LISA parameter
estimation~\cite{lisape} and in previous work on using LISA for model
selection~\cite{plowmanSMBH2}. The accretion model primarily leads to different expectations for the black hole spins (intermediate-high, $a \sim 0.6-0.9$, in the coherent case; low, $a < 0.2$, in the chaotic case). In this work we ignore black hole spin, but the accretion prescription also leaves an imprint on the component masses. The models assume that the mass-to-energy conversion efficiency, $\epsilon$, depends on black hole spin only, so the two models predict different average efficiencies of $\sim 20\%$ and $\sim 10\%$ respectively. The mass-to-energy conversion directly affects mass growth, with high efficiency implying slow growth, since for a black hole accreting at the Eddington rate, the black hole mass increases with time as
\begin{equation}
M(t)=M(0)\,\exp\left(\frac{1-\epsilon}{\epsilon} \frac{t}{t_{\rm Edd}}\right)
\end{equation}
where $t_{\rm Edd}=0.45$Gyr. The ``coherent'' versus ``chaotic'' models thus allow us to study how different growth rates affect LISA observations.

\begin{table}
\begin{tabular}{|c|c|c|c|}
\hline
Model&Seed mass prescription&Accretion prescription&LISA events per year\\\hline
SE&VHM (light seeds)&coherent&
37 (18)\\
SC&VHM (light seeds)&chaotic&
40 (21)\\\hline
LE&BVR (heavy seeds)&coherent&
24 (22)\\
LC&BVR (heavy seeds)&chaotic&
21 (18)\\\hline
\end{tabular}
\caption{\label{modelsumm}Description of the four models used in this
  analysis. The last column gives the expected number of events observed by
  LISA in one year. These were computed under the optimistic assumptions about the detector, (iii), described in the text, and the bracketed numbers were computed under the most pessimistic assumptions about the detector, (ii) in the text. While the SNRs in this paper were computed using non-spinning waveform templates, the numbers here agree well with those quoted in~\cite{lisape}, which for spinning waveform templates including higher harmonic corrections.}
\end{table}

We will again use an SNR cut, $\rho_{\rm thresh}$, to characterise whether an
event is detectable or not, and we will make both optimistic and pessimistic
assumptions about LISA in terms of the number of data streams that
are available for data analysis. At low frequency two independent data streams can be constructed from the data stream of a single LISA constellation, but one of these data streams could be lost if there is a failure on one of the three satellites in the constellation. We will therefore consider four possible scenarios: (i) 1 independent data stream, $\rho_{\rm thresh}=8$; (ii) 1 independent data
stream, $\rho_{\rm thresh}=20$; (iii) 2 independent data streams, $\rho_{\rm
  thresh}=8$; (iv) 2 independent data streams, $\rho_{\rm
  thresh}=20$. Scenarios (ii)/(iii) are the most
pessimistic/optimistic. Additionally, we include only systems that merge
within the LISA observation window in the analysis, and we consider five
different possible lengths of the LISA data set used in the analysis (3 months, 6 months, 1 year,
18 months and 2 years). It is unlikely that LISA will only take data for a few months if it works at all, but these results illustrate what we will be able to say after 3 months of observation, after 6 months and so on.

To carry out model selection, we must compute the likelihood of the data under
the various models, as given by Eq.~(\ref{like}). For the current analysis, we
use only three parameters to characterise each system: the total mass, $M$,
the mass ratio, $q$, and the source redshift, $z$. We compute the expected
rate of {\it observed} events in each bin accounting for errors in the
parameter estimation as described in Section~\ref{methods}. We compute
parameter estimation errors using the Fisher matrix approximation for
quasicircular, non-spinning BH binary inspirals modelled in the restricted
post-Newtonian approximation, following~\cite{Berti:2005qd}. Cosmological
parameter uncertainties and weak lensing errors are folded in as described
above. The fact that we are ignoring spins, eccentricity and higher harmonics
in our analysis has two consequences. Firstly, our estimates for the
signal-to-noise ratio for each source are pessimistic, because spins and
higher harmonics of the signal (which are neglected in our model) usually
increase the energy radiated and the mass reach of the detector. Secondly,
spin encodes important information about the history of a particular BH, and
in particular its accretion history \cite{bv}. This could help resolve models
where BHs are ``born equal'' but grow via different mechanisms. In this sense
our results should be considered conservative.

\begin{figure}
\begin{tabular}{cc}
\includegraphics[width=0.5\textwidth]{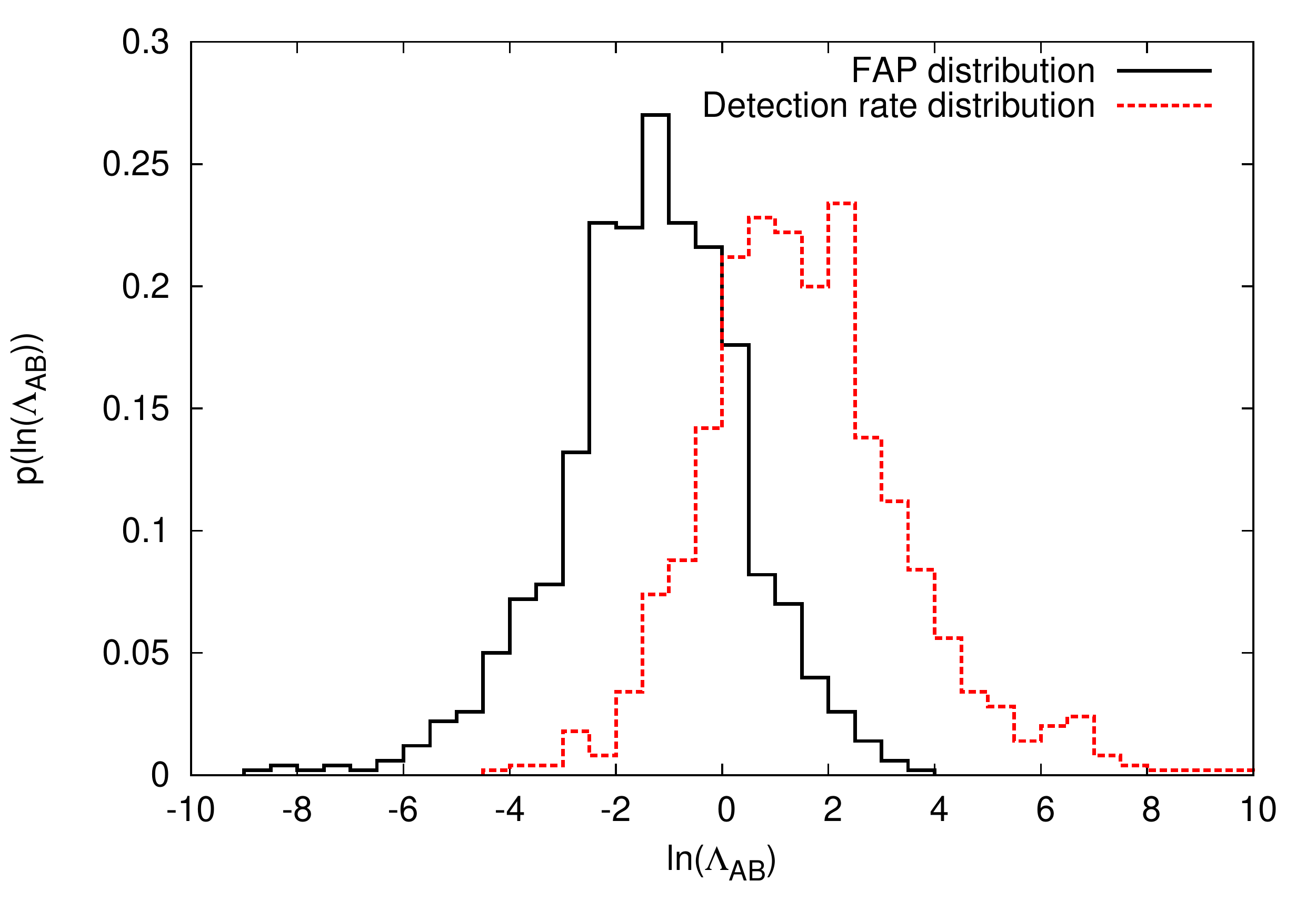}&
\includegraphics[width=0.5\textwidth]{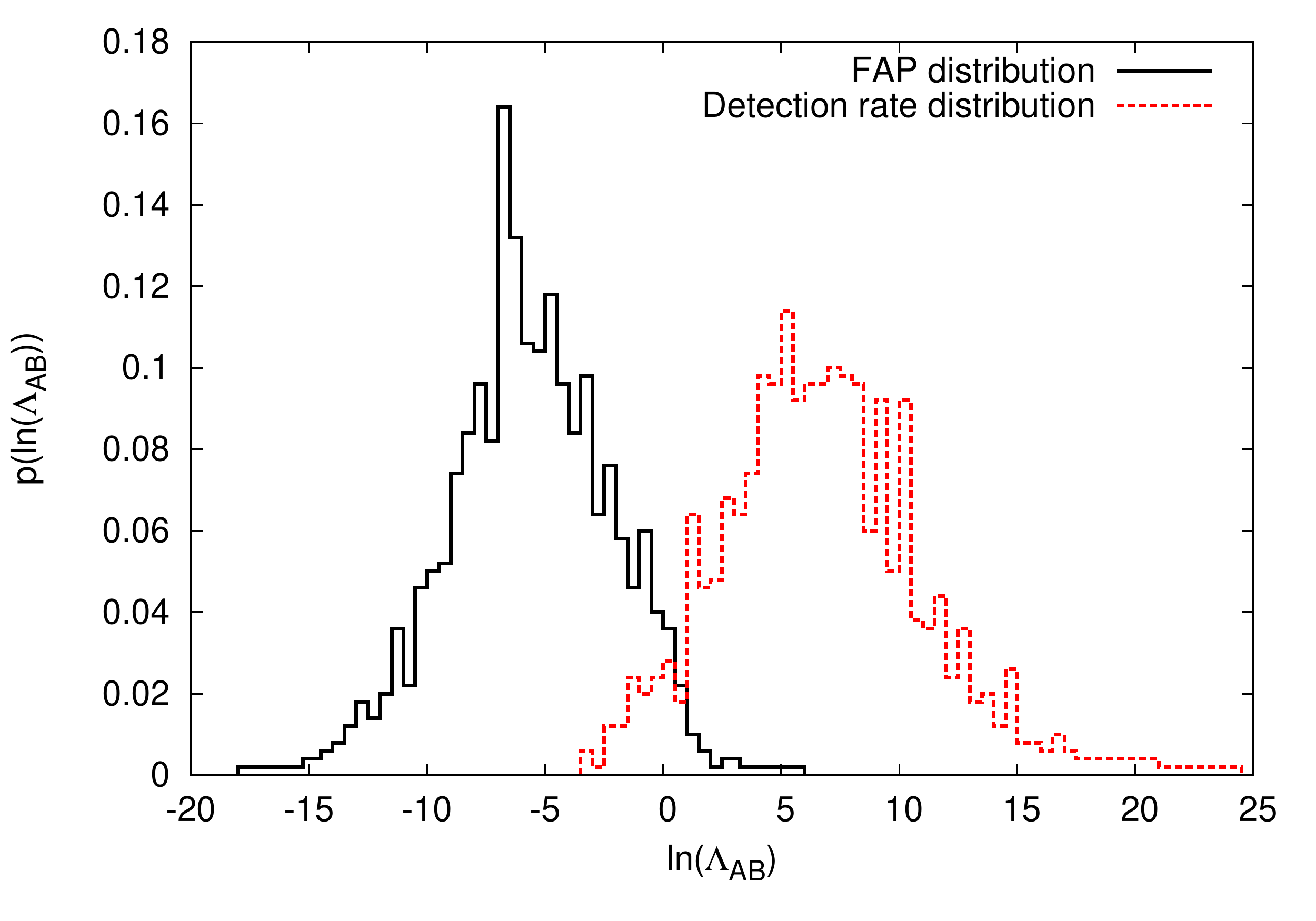}
\end{tabular}
\caption{\label{likedistfig}We show the distribution of the logarithm of the likelihood ratio,
  $\ln(\Lambda_{AB})$, for comparison of model SE to model SC using 3 months of LISA
  data (left panel) and using 1 year of data (right panel). The
  curves labelled ``FAP distribution'', where FAP stands for ``false alarm
  probability'', describe realisations drawn from the ``wrong'' model (SC).
  Those labelled ``detection rate distribution'' describe realisations
  drawn from the ``right'' model (SE). We note that the distribution for one year of data is somewhat broader than that for 3 months. This is not unsurprising, since more events are expected in one year of data, and both the mean and the variance of the underlying Poisson distribution increase with the total number of events.}
\end{figure}

The four models we use do not have free parameters, so we cannot determine a
posterior probability distribution on the model parameters. Instead, we can
use Eq.~(\ref{odds}) to decide which model provides the best description of the
data. The models we are comparing have all been constructed to be consistent with existing constraints from observations in the electromagnetic spectrum, so there is presently no reason to prefer one model over the others. We therefore assume equal prior probabilities on all models, so the
odds ratio reduces to the likelihood ratio
\begin{equation}
\Lambda_{AB} = \frac{p(D|M=A)}{p(D|M=B)} .
\end{equation} 
It is clear that when $\Lambda_{AB} \gg 1$, model $A$ should be
preferred. What value of $\Lambda_{AB}$ is sufficient to make such a
statement? This can be answered by looking at the distribution of
$\Lambda_{AB}$ over many realisations of model $A$ and model $B$. We generate
a realisation of model $X$ by drawing events from the underlying population,
applying the appropriate SNR cut, and adding in parameter errors to each
event. We can then compute the likelihoods for this event set under model $A$
and model $B$ and hence $\Lambda_{AB}$. In Figure~\ref{likedistfig} we show
the distribution of $\ln(\Lambda_{AB})$ in 1000 realisations each of model $A$ and
model $B$.

\begin{figure}
\begin{center}
\includegraphics[width=0.75\textwidth]{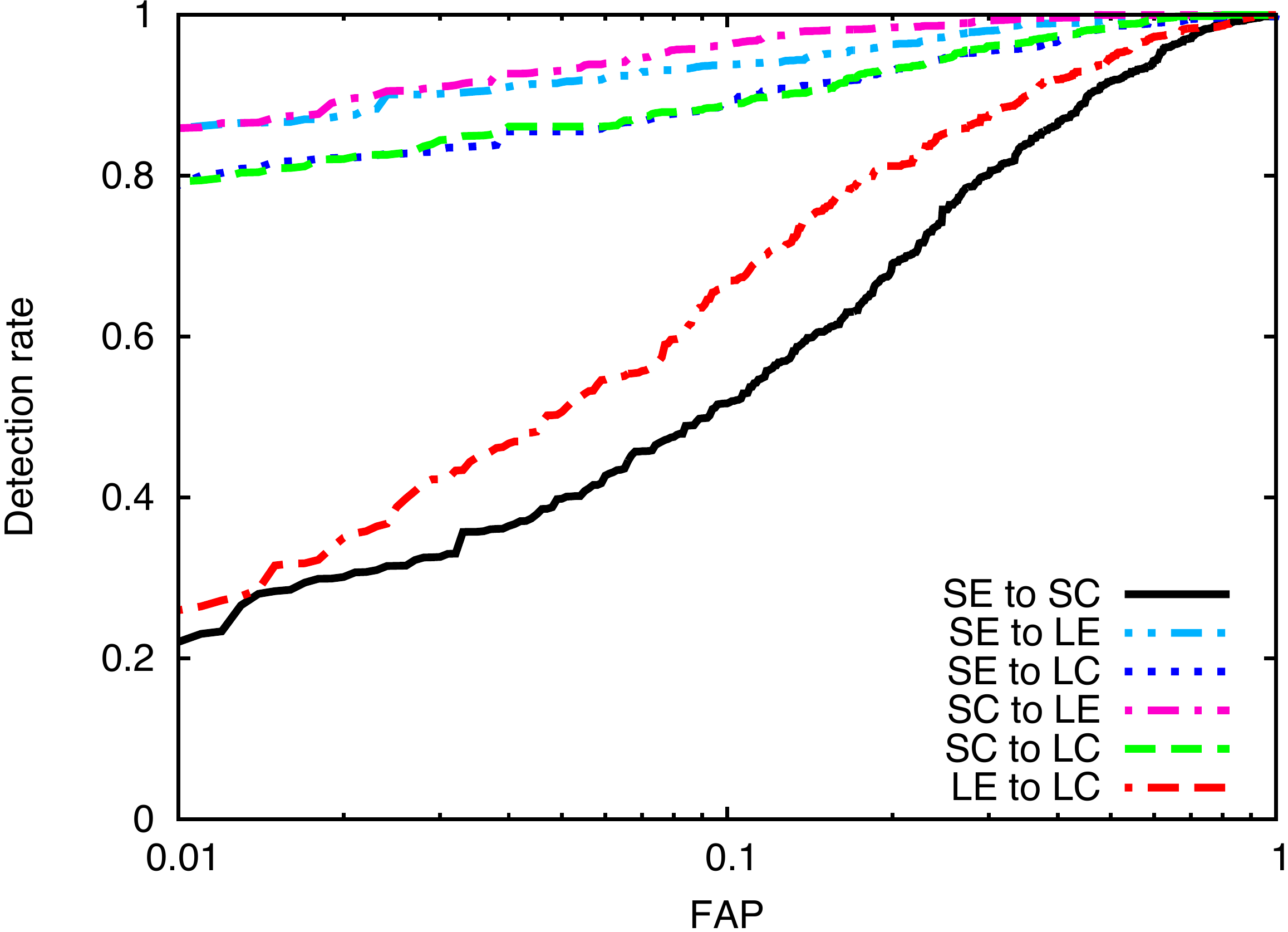}
\end{center}
\caption{\label{ROCallcomp}Receiver operator characteristic curves for
  different model comparisons. These were computed as described in the text
  for a fixed LISA mission duration of 3 months and assuming detector scenario
  (ii), i.e., one data channel only and $\rho_{\rm thresh}=20$.}
\end{figure}

As we would expect, when the realisation is drawn from model $A$,
$\Lambda_{AB}$ tends to be greater than $1$, while when it is drawn from
model $B$, $\Lambda_{AB}$ tends to be less than $1$. For a given choice of
threshold on $\Lambda_{AB}$, points in the model $A$ histogram to the right of
that threshold represent ``detections'', i.e., realisations in which model $A$
would be chosen over model $B$ when model $A$ was correct. Points in the model
$B$ histogram are ``false alarms'', i.e., realisations in which model $A$ is
chosen over model $B$ when in fact model $B$ is correct. The histograms become
better separated when using a longer segment of LISA data, since we have more events
in that case. Another way to represent this information is through a
receiver-operator-characteristic (ROC) curve, which shows ``detection''
probability versus ``false alarm'' probability (FAP). For a given threshold on
$\Lambda_{AB}$, the detection probability is the fraction of realisations of
model $A$ that lie to the right of that threshold, while the false alarm
probability is the fraction of realisations of model $B$ that lie to the
right. In Figure~\ref{ROCallcomp}, we show the ROC curves for all possible
comparisons between the four models, for a 3 months of LISA data  and with
the most pessimistic scenario, (ii), for the detector performance. The ROC
curve is a frequentist way to represent the performance of an algorithm, but
it encodes similar information to the Bayesian approach of assigning
probabilities of $p(D|M=A)/[p(D|M=A)+p(D|M=B)]$ and
$p(D|M=B)/[p(D|M=A)+p(D|M=B)]$ to models $A$ and $B$ respectively. Note that
the false alarm and detection rates are {\em per LISA observation}: a FAP of $0.1$
indicates that if a LISA observation of this duration was repeated independently $10$ times, we
would expect to incorrectly choose model $A$ only once.

\begin{figure}
\includegraphics[width=0.5\textwidth]{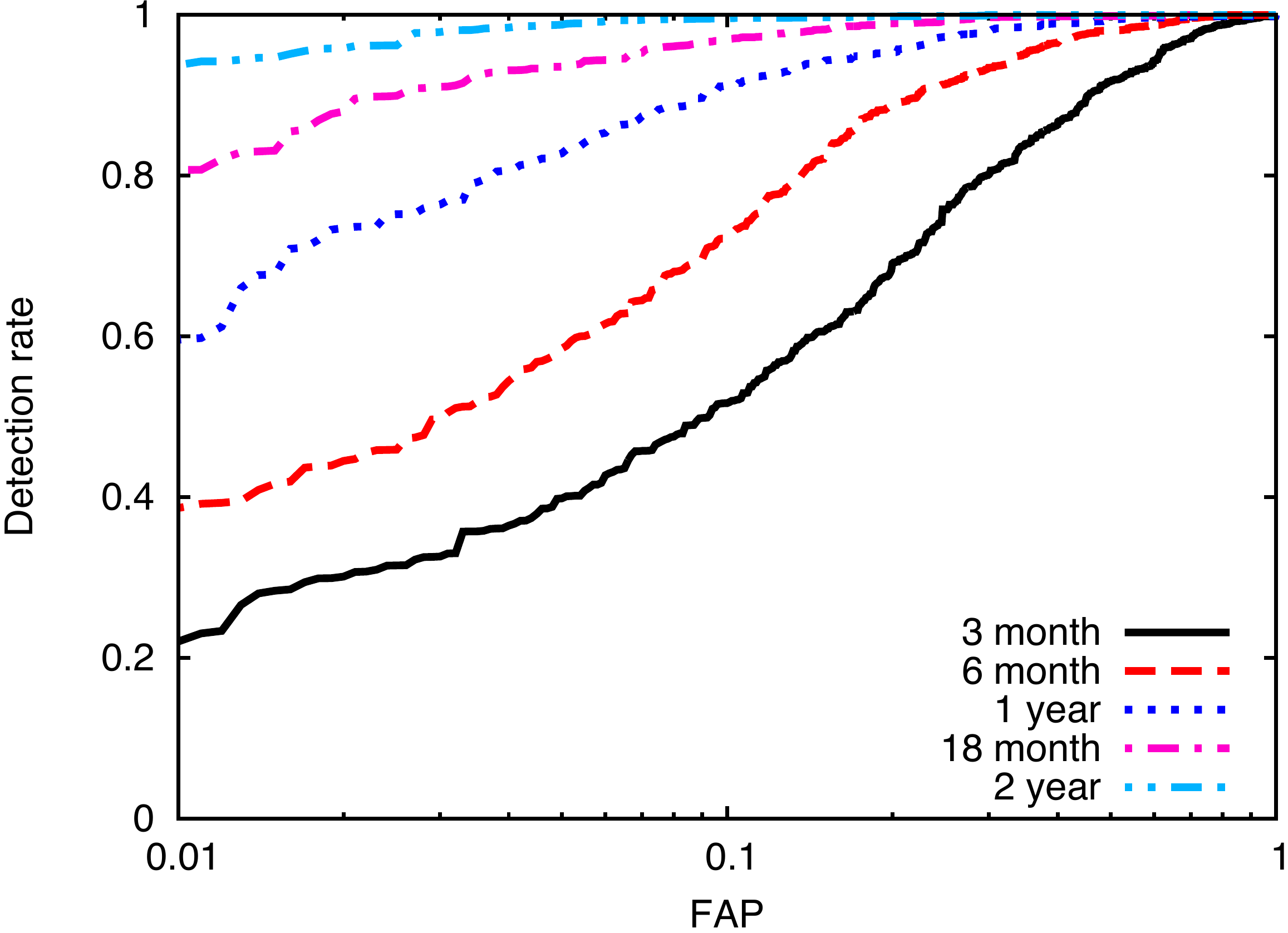}
\includegraphics[width=0.5\textwidth]{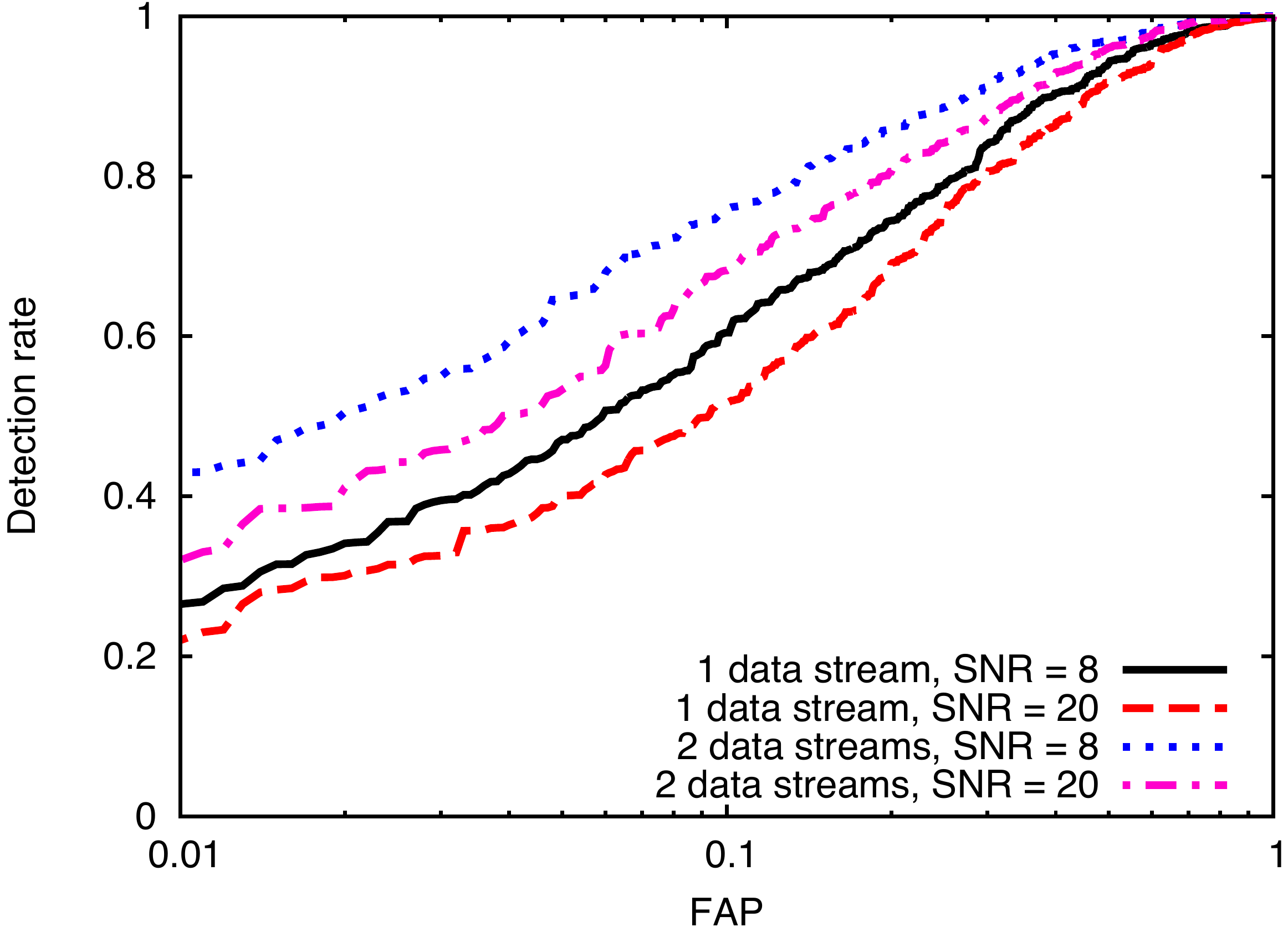}
\caption{\label{ROCallscen} As Figure~\ref{ROCallcomp} but now for a fixed model comparison (SE versus SC), but varying the length of the LISA observation for fixed detector scenario (ii) (left) and varying the detector scenario for a fixed LISA observation of 3 months (right).}
\end{figure}

It is clear from Figure~\ref{ROCallcomp} that even using as
little as 3 months of data, LISA can easily choose between the four
models. With the exception of the SE/SC and LE/LC comparisons, each pair of
models shows a detection rate in excess of $80\%$ at a $1\%$ false alarm
rate. Models SE and SC are most difficult to distinguish with a detection rate
of only $50\%$ at a $10\%$ false alarm rate, followed by models LE and LC with
a detection rate of $60\%$ at that FAP. This is to be expected, as these pairs
of models have the same seed mass distribution and differ only in the
accretion prescription, so the distribution of {\it masses} for the events are
quite similar (this pessimistic conclusion would most likely change if we
included spins in our model waveforms). In the other cases, the mass
distributions are quite distinct so the accurate mass measurements that are
possible with LISA allow discrimination of the models with only a handful of
events. In the left panel of Figure~\ref{ROCallscen} we consider the SE to SC
comparison and detector scenario (ii) only and show how the ROC performance
depends on the mission duration. We see that our ability to distinguish models
increases rapidly with the duration of the observation. For a 1 year observation we can
distinguish SE and SC with a rate of $\sim80\%$ for an FAP of $5\%$, which is
comparable to what can be achieved using a 3 month observation for the other comparisons.
%

In the right panel of Figure~\ref{ROCallscen} we again restrict to the SE to SC comparison, but
now fix the observation to 3 months and show how the performance depends
on the detector scenario. There is a relatively modest increase in performance
for the more optimistic detector scenarios. At an FAP of $10\%$, the detection
rate increases from $\sim50\%$ to $\sim70\%$ going from the most pessimistic
to the most optimistic scenario. The detector performance has a relatively
weak effect since many of the events with the greatest distinguishing power
have very high SNR and can be seen under any scenario. The use of two data streams rather than just one both increases the SNR of an event with given parameters and reduces the parameter estimation errors that arise due to instrumental noise. It is clear from Figure~\ref{ROCallscen} that both of these are important, since both of the 2 data stream curves lies above both of the 1 data stream curves. If the SNR increase alone was important, we would expect the 2 data stream, SNR $=20$ curve to lie between the one data stream curves, as SNR $=20$ in two data streams corresponds to SNR $\approx 14$ in each data stream.

Reference~\cite{plowmanSMBH2} also studied MBHM model selection with LISA
considering the same four models using the non-parametric Kolmogorov-Smirnov
test to compare the distributions of one or two parameters between
models. Their conclusions were broadly the same, i.e., that LISA can tell
between models very easily. The approach described here has several advantages
over theirs: we have a parametric model using the reasonable assumption of a
Poisson distribution in each bin; we use the distribution of {\it all} model
parameters simultaneously to compare models (3 in this case, but it is easily
extendable to more); the present framework naturally extends to parameterised
models for the underlying BH population, as used in the EMRI case; and, as
described in Section~\ref{methods}, once actual LISA data is available, we
will be able to fold the measured uncertainties in the source parameters into
the analysis, rather than relying on theoretical estimates.

\section{Discussion} 
\label{discuss}
We have described a framework for using the set of events observed by LISA to
constrain models of the massive BH population in the Universe. Using
EMRI events, we should be able to constrain the slope of a parametric model
for the mass function of BHs in the LISA range to a precision of $\sim
\pm0.2$ with just ten observed events and this improves with the number of
observed events as $N_{\rm obs}^{-1/2}$. EMRI events alone will not be able to
constrain any evolution of this mass function with redshift. LISA MBHM events
can be used to choose between different models for the assembly of structure
in the Universe. Assuming that the LISA events were drawn from one of four
simple models, we have shown that we would be able to confidently identify the
correct model after collecting as little as three months of LISA data. BHs in the
LISA mass range, $10^4M_\odot$--$10^7M_{\odot}$, are not well constrained by
current data, and so LISA has the potential to significantly improve our
understanding of such systems and how they were assembled.

These preliminary results could be extended in several ways. The parameter
space considered in the EMRI case should be expanded to explore what we can
learn from EMRI measurements of BH spins and the eccentricities, inclinations
etc. of EMRI orbits. MBHM waveform models including the merger/ringdown signal
\cite{BCW} and spin precession dynamics should improve our ability to
distinguish between models.  It will also be important to explore whether LISA
EMRI and low-$z$ MBHM observations {\em together} can probe the redshift
evolution of the mass function and break the degeneracy between the mass
function and the mass-scaling of the EMRI rate. 

In the MBHM model selection context, the four models considered were
deliberately chosen in~\cite{lisape} to be as different as possible in order
to provide ranges for the parameter estimation accuracies that might be
achieved by LISA. It is therefore perhaps unsurprising that LISA will be able
to distinguish these models very easily. The real Universe is likely to be a
hybrid between light and heavy seed models, so it will be informative to
explore more realistic mixtures between the present (oversimplified) seeding
and accretion prescriptions. Indeed, if parameters can be introduced into the
models that characterise the input physics (e.g., the seed mass distribution
and accretion efficiency), we could determine the precision with which LISA
will be able to measure such parameters. The results described here provide a
clear illustration of the significant astrophysics that can be done with LISA
observations and indicate that it will be worthwhile to carry out such more
sophisticated analyses.

\ack JG's work is supported by the Royal Society. EB's research was supported
by NSF grant PHY-0900735. MV was supported by NASA ATP Grant NNX07AH22G and a Rackham faculty grant.

\section*{References}
\bibliography{ModelSelLISA8}

\end{document}